\documentclass[twocolumn,prc,floatfix,showpacs,preprintnumbers,nofootinbib,%
superscriptaddress]{revtex4}
\usepackage[utf8x]{inputenc}
\usepackage{mathrsfs}
\usepackage{amssymb}
\usepackage{amsmath}
\usepackage{graphicx}
\usepackage{ulem}

\usepackage{xcolor}
\definecolor{lcolor}{rgb}{0.5,0,0}
\definecolor{citcolor}{rgb}{0,0.3,0.0}

\usepackage[breaklinks,colorlinks,urlcolor=blue,citecolor=citcolor,linkcolor=lcolor]{hyperref}
\usepackage{mciteplus}

\newcommand{\tev}{\ \textrm{TeV}}
\newcommand{\mb}{\ \textrm{mb}}
\newcommand{\fm}{\ \textrm{fm}}

\def\min{{\text{min}}}

\begin{document}

\author{Wei-Tian Deng}
\affiliation{
School of Physics, Huazhong University of Science and Technology, Wuhan 430074, China\\
}

\author{Hirotsugu Fujii}
\affiliation{
Institute of Physics,
University of Tokyo, Tokyo 153-8902, Japan
}

\author{Kazunori Itakura}
\affiliation{
KEK Theory Center, IPNS, KEK,
Tsukuba 305-0801, Japan}
\affiliation{
Department of Particle and Nuclear Studies, Graduate University for 
Advanced Studies(SOKENDAI),
Tsukuba 305-0801, Japan}

\author{Yasushi Nara}
\affiliation{
Akita International University, Yuwa, Akita-city 010-1292, Japan
}

\title{
Forward Hadron Productions in high energy pp collisions
from a Monte-Carlo generator for Color Glass Condensate
}


\preprint{}

\begin{abstract}
We develop a Monte-Carlo event generator based on 
combination of a parton production formula including the 
effects of parton saturation (called the DHJ formula) and 
hadronization process due to the Lund string fragmentation model. 
This event generator is designed for the description of 
hadron productions at forward rapidities and in a wide transverse 
momentum range in high-energy proton-proton collisions. 
We analyze transverse momentum spectra of charged hadrons 
as well as identified particles; pion, kaon, (anti-)proton 
at RHIC energy, and ultra-forward neutral pion spectra from 
LHCf experiment. 
We compare our results to those obtained in other models 
based on parton-hadron duality and fragmentation functions.
\end{abstract}

\maketitle

\section{Introduction}

Theoretical understanding of hadron production in high energy 
hadronic interactions is one of the most relevant subjects 
in QCD. For example, 
in ultra-relativistic heavy-ion collisions,
the subsequent space-time evolution of a quark-gluon plasma (QGP)
is described by relativistic hydrodynamics~\cite{Heinz:2013th}.
Thus it is very important to fix a reliable initial condition
for hydrodynamical simulations in order to extract the correct
dynamics of the reaction. 
Another example is the air-shower development induced by
ultrahigh-energy cosmic-rays (UHECR), which is
very sensitive to high-energy hadronic interactions~\cite{Engel:2011zzb}.
The interactions are usually simulated using a Monte-Carlo 
event generator, and the choice of an interaction model 
may critically affect the air shower analysis.

As an incident energy increases in a hadronic collision,
reactions involving small Bjorken's-$x$ gluons
become dominant, and a framework which
treats a highly dense gluonic system is needed.
The Color Glass Condensate (CGC) framework has been
proposed in order to describe such a dense gluon system~\cite{MV,GIJV}
in which the saturation scale $Q_s$ characterizes 
nonlinear nature of the system.

For practical computation of gluon production in the CGC framework,
two different approaches are often taken:
solving classical Yang-Mills equation
on the lattice~\cite{KV,Lappi,IPGlasma}
or the $k_{T}$-factorized production formula 
with unintegrated gluon distribution (uGD) with saturation~\cite{ktfact}.
They have been successful in explaining many experimental data
at RHIC and LHC energies.
The models based on the $k_{T}$-factorization formula,
such as KLN model given in Ref.~\cite{KLN} 
and its extensions~\cite{fKLN,Levin:2010zy,Levin:2010dw,Levin:2011hr},
describe hadron multiplicity distributions
and transverse momentum ($p_T$) distributions in low momentum region
under the assumption of Local Parton-Hadron Duality (LPHD)~\cite{LPHD}.
On the other hand, in high momentum region, it can be well 
reproduced by the $k_{T}$-factorization formula
combined with the fragmentation 
functions~\cite{Kharzeev:2004yx,Albacete:2010bs,
Fujii:2011fh,Fujii:2012zza,Albacete:2012xq,Lappi:2013zma}.

A hybrid formalism was proposed in Ref.~\cite{DHJ}
to describe the collisions between a
dilute projectile and a dense target 
(Dumitru-Hayashigaki-Jalilian-Marian (DHJ) formula).
It has been applied to the computations of forward 
particle production spectra,
not only for charged hadrons and pions,
but also for baryons~\cite{Hayashigaki:2006ek} in 
d+Au collisions.
The DHJ formula was also utilized to compute
baryon productions for Au+Au collisions
in Ref.~\cite{MehtarTani:2009dv,Duraes:2014jxa},
and it is found that transverse momentum spectra and
net-baryon rapidity distributions in Au+Au collisions
at RHIC are well described with the DHJ formula.

Substantial progresses were made recently:
the Balitsky-Kovchegov equation with running-coupling accuracy 
(rcBK equation)~\cite{rcBKkern} was obtained, and numerical 
methods have been developed to solve the rcBK evolution~\cite{AK}.
A global analysis was performed 
with the rcBK equation for the nucleon structure function 
measured at HERA at small values of $x\leq 0.01$,
which yields good fits (the AAMQS parametrization)~\cite{AAMQS}.
At the same time, when applied to hadronic interactions,
the AAMQS parametrization
provides a good agreement with the data at RHIC and LHC
\cite{Albacete:2010bs,Fujii:2011fh,Fujii:2012zza,Albacete:2012xq,
Lappi:2013zma}.
For a recent review, see Ref.~\cite{Albacete:2013tpa}.
The next-to-leading order (NLO) corrections to the hybrid formula
were computed in Refs.~\cite{NLODHJ}.
It is found that the NLO corrections yield the same $p_T$ 
dependence as the leading-order (LO) expression in the regime 
$p_T\leq Q_s$, where non-linear effects are strong,
and thus the LO formula can be applied with a constant $K$-factor
at forward rapidities.

Geometrical fluctuations of the projectile and the target 
are important in nucleus-nucleus collisions. They are included 
in the $k_T$-factorization approach within a Monte-Carlo 
based formulation~\cite{Albacete:2012xq,MCKLN,MCKT}
and, this approach has been extensively
used as initial conditions for the 
subsequent hydrodynamical evolution 
of a system~\cite{Hirano:2012kj,Hirano:2010jg,Qiu:2011hf}.
However, it only provides the (energy) density distribution at each grid.
Full event generation of all particles with their 4-momenta assigned
has not yet been implemented along this approach.

The first attempt to generate full parton configurations based on
the KLN $k_T$ factorization formula 
was done in Ref.~\cite{Drescher:2004sd} (BBL Monte-Carlo model),
and it was applied to high-energy Cosmic Ray air shower simulations.
In BBL, momenta of quarks and gluons are generated according to
the $k_T$-factorization formula, and 
the hadronization is performed using 
the Lund string fragmentation model.
This approach allows one to describe particle production
from low to high momentum region consistently.
They showed that atmospheric air showers are sensitive to the
interactions with partons at very small $x$.

In this paper, 
we present a newly developed Monte-Carlo event generator
based on the DHJ formula which implements
the latest theoretical update of
the uGD function
from the numerical solution of the rcBK  equation.
Specifically, we generate partons according to the DHJ formula together with
initial and final state radiations based on the DGLAP evolution equation.
Strings are formed by those produced partons and remnants
which are fragmented into hadrons by the Lund string fragmentation model.
We will compare our results to the forward hadron spectra
in proton-proton collisions observed at RHIC and LHCf experiments,
and discuss the mechanism of the particle production.
Results with the LPHD and fragmentation function adopted to the DHJ formula
are shown for comparison with our Monte-Carlo approach.

This paper is organized as follows.
In Sec.~\ref{sec:model}, implementation of the
DHJ formula into the Monte-Carlo generator is explained.
In Sec.~\ref{sec:results}, we compare our numerical results to
the transverse momentum distribution in proton-proton collisions
at RHIC and LHC in forward rapidity regions.
Results from different approaches are also shown for 
comparison.
A summary is given in Sec.~\ref{sec:summary}.

\section{The DHJ+Lund Model}\label{sec:model}

We consider high-energy proton-proton scatterings at forward 
rapidities.
We shall employ the DHJ hybrid formalism 
where we treat the collision between a large-$x_1$ 
($=(p_T/\sqrt{s})\exp(y)$)
parton (quarks or gluons) from the projectile and a small-$x_2$ 
($=(p_T/\sqrt{s})\exp(-y)$) 
gluon from the target. Then, the forward parton 
production cross-section with transverse momentum $p_T$ and rapidity $y$
is given as
\begin{equation}
\frac{d\sigma_\text{DHJ}}{dyd^2p_{T}}
  =\frac{K}{(2\pi)^2}\frac{\sigma_0}{2}\sum_{i=q,g}
   x_1 f_{i/p}(x_1,Q^2)N_i(x_2,p_T)
,
  \label{eq:dhj}
\end{equation}
where $f_{i/p}$ is the collinear parton distribution function (PDF) for
a large-$x_1$ parton $i$, 
$N_{i}$ ($i=F,A$) is the Fourier transform of the dipole 
scattering amplitude in the fundamental (for quark or anti-quark 
scattering)  or adjoint (for gluon scattering) representation
for small-$x_2$ gluon. 
We use $Q=p_T$ as a factorization scale for PDF as a default value
for the DHJ+Lund model.
As a default setting in our model, CTEQ5L~\cite{Lai:1999wy} for PDF and
$K$-factor of $K=1.0$ are used.
The average transverse area of the proton $\sigma_0/2=16.5\mb$
is obtained by the DIS fits at HERA~\cite{AAMQS,Lappi:2013zma}.
The uGD functions $N_{F,A}$
are obtained as the numerical solution of the rcBK equation 
and are fitted to the HERA data~\cite{AAMQS}.
Specifically, parameter set g1.101 in Ref.\cite{Albacete:2012xq}
is used in this paper in which the initial condition for $N_F$ in 
the coordinate representation at $x_0=0.01$ is taken to be
\begin{equation}
N_F(r,x_0) = 1 - \exp\left[ -\frac{(r^2Q_{s0}^2)^\gamma}{4}
          \ln\left(\frac{1}{\Lambda r}+e\right) \right]\ ,
\end{equation}
where $r$ is the transverse size of a color dipole,
$\gamma=1.101$, $Q_{s0}^2=0.157$ GeV$^2$, $\Lambda=0.241$ GeV,
and we assume impact parameter independent rcBK equation.

\begin{figure}[tbp]
\centerline{\includegraphics[width=0.45\textwidth]{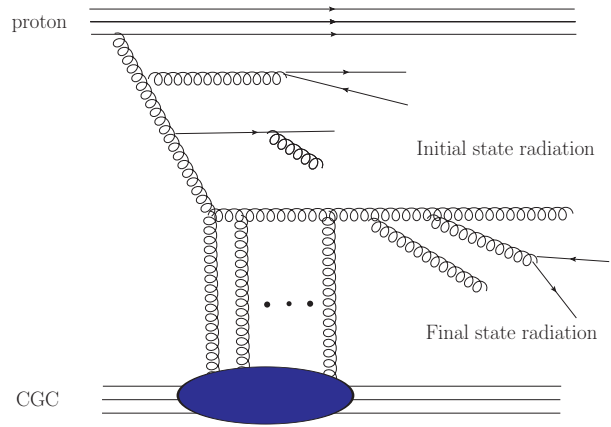}}
\caption{Schematic picture of a possible event
in which a interaction $gg\to g$ with initial and final
state radiations is simulated by the DHJ+Lund model.
\label{fig:mcdhj}
}
\end{figure}

In Fig.~\ref{fig:mcdhj}, a schematic picture of what the DHJ+Lund model
simulates is illustrated for $gg\to g$ together with initial and final
state radiations.
Gluons and quarks are generated randomly according to the formula
Eq.(\ref{eq:dhj}) with the minimum momentum $p_{T,\min}=1$ GeV.
We have checked that the transverse momentum distributions
is insensitive to $p_{T,\min}$: Simulations with $p_{T,\min}=0.25$ GeV yields
similar results as
$p_{T,\min}=1$ GeV which will be presented in this paper.
Initial  radiations for high $x$ projectile part and final state radiations
are included in line with PYTHIA approach~\cite{pythia6,pythia}.
But explicit gluon productions from the initial state radiation 
due to $x$-evolution is not included in this work.

The effects of multiple parton interaction (MPI) is simulated based on the
eikonal model~\cite{Eikonal},
which is used in several event generators 
such as HIJING~\cite{Wang:1991hta,Gyulassy:1994ew,Deng:2010mv},
SIBYLL~\cite{Fletcher:1994bd,Ahn:2009wx},
and HERWIG++~\cite{Bahr:2008dy,Bahr:2008pv}.
By assuming that each interaction is independent, the probability to
have $n$ scattering is given by
\begin{equation}
 P_n = \frac{n(b,s)^n}{n!}\exp(-n(b,s))
,
\end{equation}
where
\begin{equation}
  n(b,s)= A(b) \sigma_\text{DHJ}(s)
\end{equation}
is the average number of partonic collisions at a given impact parameter $b$
with the invariant mass of the collision $s$.
$A(b)$ is the spatial overlap of the two colliding hadrons, and
we take the Gaussian form:
\begin{equation}
A(b)=\frac{1}{4\pi B}\exp\left(-\frac{b^2}{4B}\right)
\end{equation}
with the parameter $B=0.25\fm^2$.
Notice that the multiple scattering of an incoming parton with 
a coherent color field in the target CGC is already included in 
$\sigma_{\rm DHJ}$. MPI describes events which include several 
hard scatterings.
In an event with several interactions,
we will have several hard scatterings of the types $gg\to g$
or $qg\to q$. Among several hard scatterings,
quark production from $qg\to q$ process is generated only once,
if it is selected,
and the rest of all interactions 
are assumed to be $gg\to g$ scattering for simplicity.
More sophisticated implementation of multiple parton interaction
will be discussed elsewhere.

We need to introduce hadronization of parton in order to
compare with experimentally measured hadrons.
Hadronization process is entirely nonperturbative process and
we have only phenomenological approaches.
The hypothesis of LPHD~\cite{LPHD}
has been formulated
based on the observation that
parton distribution computed in perturbative QCD (pQCD) approach
gives good description of hadrons even at small $p_T$.
On the other hand, the fragmentation function
is used to hadronize the partonic system above the factorization scale.
The Lund string model of hadronization has been 
developed based on the massless relativistic string as a model
for QCD color field~\cite{Andersson:1983ia},
and implemented in the Monte Carlo event generator PYTHIA~\cite{pythia6}.
In this paper, we utilize the Lund model for the hadronization.

\begin{figure}[tbp]
\centerline{\includegraphics[width=0.3\textwidth]{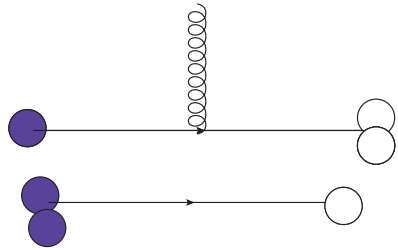}}
\caption{Schematic picture of a possible string configuration 
after the collision, where two strings are formed and stretched
 between quarks and diquarks.
The gluon which is produced in the $gg\to g$ process is attached to
one of the strings.
\label{fig:string}
}
\end{figure}

Besides hard scatterings between the two hadrons,
two strings are generated with
the fractional energy $x$ of the quark,
which is chosen according to the probability profile:
\begin{equation}
P(x) = \frac{(1-x)^\alpha}{\sqrt[4]{x^2+c/s}}
\label{eq:dpm}
\end{equation}
with the default parameter in PYTHIA6~\cite{pythia6}:
$\alpha=3$ and $c=0.36$.
We use PYTHIA 8.186~\cite{pythia8} to simulate 
the string fragmentation into hadrons.
A simplest string configuration
in a Monte-Carlo event in our model is depicted in Fig.~\ref{fig:string}.
When one $gg\to g$ interaction
occurs and there is no radiation, then 
there are two strings formed, one of which
will have one gluon attached.
There are of course many other string configurations in the simulation, 
although we do not show all of them here.

\section{Comparison with the experimental data}
\label{sec:results}

In this section, we compare our results to the experimental data
at both RHIC and LHC energies.
Within the DHJ+Lund model, we have generated 70 million events 
at RHIC energy, and 1 million events at LHC energy, which were then 
used to compute the transverse momentum distributions of the produced 
hadrons in pp collisions.

For other model approaches,
we present the results of 
the DHJ formulation with the LPHD ansatz (DHJ+LPHD)
in low momentum region and those 
of the DHJ formulation with fragmentation function (DHJ+FF)
in high momentum region.
We also compare PYTHIA and HIJING  results in which
both soft and pQCD mini-jet productions are included.

\subsection{Charged hadrons}
\label{sec:charged}

\begin{figure}[bp]
\centerline{\includegraphics[width=0.45\textwidth]{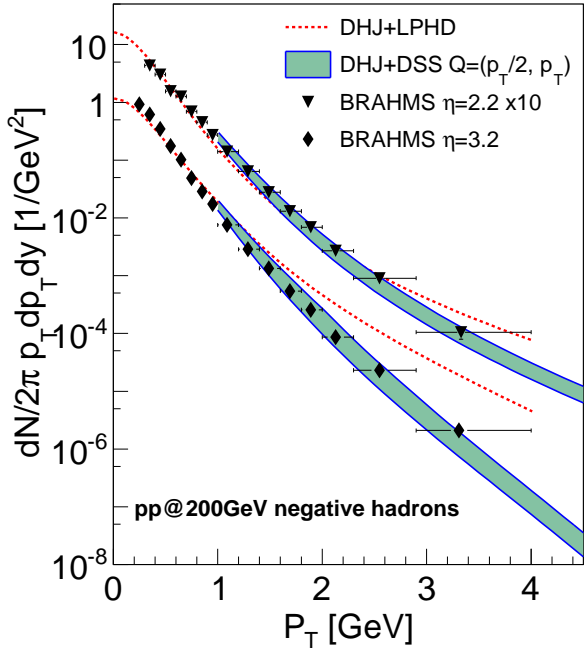}}
\caption{Negatively charged
 hadron transverse momentum distributions in pp collisions
at $\sqrt{s}=200$ GeV
at $y=2.2$ and $3.2$ from BRAHMS experiment~\cite{Arsene:2004ux}
are compared with the DHJ+LPHD model and DHJ+DSS independent fragmentation.
$K=2.5$ is used for both calculations.
In the DHJ+DSS results, scale dependence between $Q=p_T/2$ and $Q=p_T$
is shown by the width of the band.
\label{fig:lphd200}
}
\end{figure}

For later discussions, let us first explain the DHJ+LPHD model briefly.
We basically follow the same approach as in Ref.~\cite{Levin:2010dw}:
Transverse momentum of the parton $q_T$ is obtained by
$q_T=p_T/\langle z \rangle$ in Eq.~(\ref{eq:dhj})
and $q_T$ is replaced by $m_T=\sqrt{q_T^2+m^2_\text{jet}}$
in order to take into account mass effect.
In the framework of the LPHD in Ref.~\cite{Levin:2010dw}, 
$\langle z \rangle \approx 0.5$ was used 
and good agreement with the data was obtained 
for the transverse momentum distributions for charged hadrons
at low transverse momentum range less than 4 GeV
at mid-rapidity at $\sqrt{s}=2.36\tev$.
Thus in the case of forward rapidity $y$ region
we take an average of $z$ value as
\begin{equation}
\langle z \rangle= \frac{(1+z_\text{min})}{2}
    =\frac{1+\frac{m_T}{\sqrt{s}}e^y}{2}
.
\label{eq:z}
\end{equation}
The saturation scale $Q_s$ is used
for the factorization scale in the PDF.

\begin{figure}[tbp]
\centerline{\includegraphics[width=0.45\textwidth]{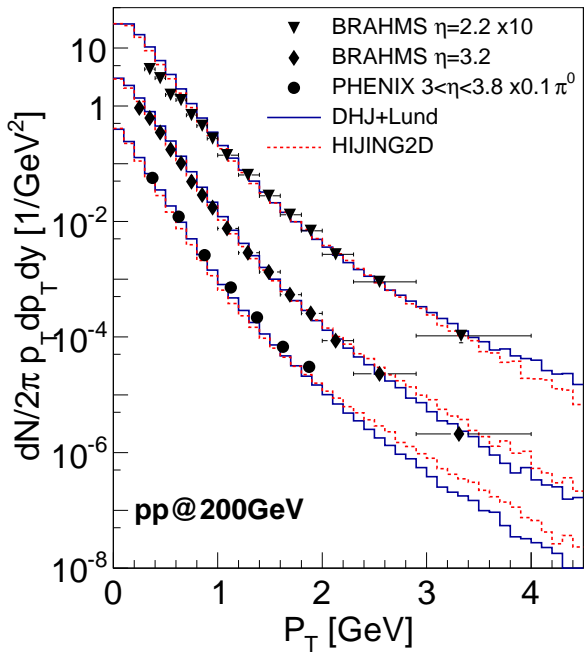}}
\caption{Negatively charged
 hadron spectra and neutral pion spectra in pp collisions
at $\sqrt{s}=200$ GeV. Data are from the BRAHMS~\cite{Arsene:2004ux}
and PHENIX~\cite{Adare:2011sc}.
Results from DHJ+Lund (solid lines) and HIJING2D (dotted lines) are
compared.
\label{fig:dhj200}
}
\end{figure}
\begin{figure}[tpb]
\centerline{\includegraphics[width=0.45\textwidth]{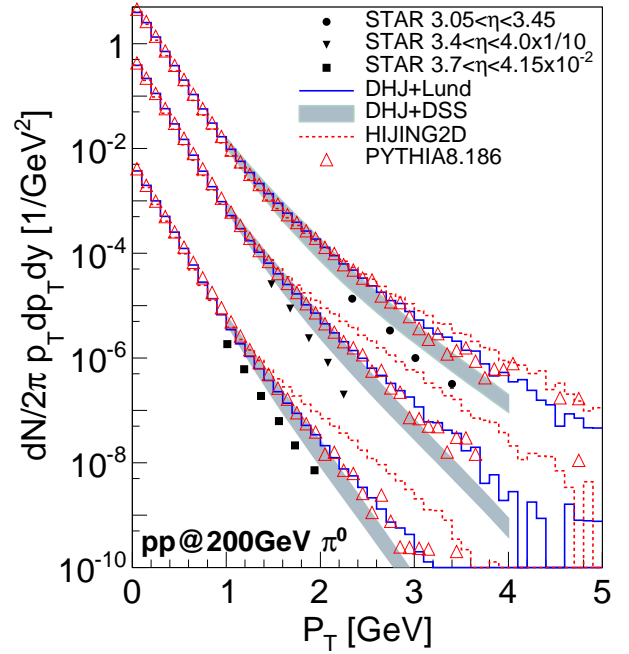}}
\caption{Neutral pion invariant cross-sections from STAR~\cite{Adams:2006uz}
in pp collisions at $\sqrt{s}=200$ GeV are compared to
the DHJ+Lund model (solid lines), HIJING2D (dotted lines),
PYTHIA8.1 (triangles), and DHJ+DSS (bands).
The DHJ+DSS results are for $K=2.5$ and $Q=(p_T/2,p_T)$.
All theoretical results are obtained by averaging over given rapidity bin.
\label{fig:starpi0}
}
\end{figure}

\begin{figure*}[t!]
\begin{center}
\includegraphics[width=0.45\textwidth]{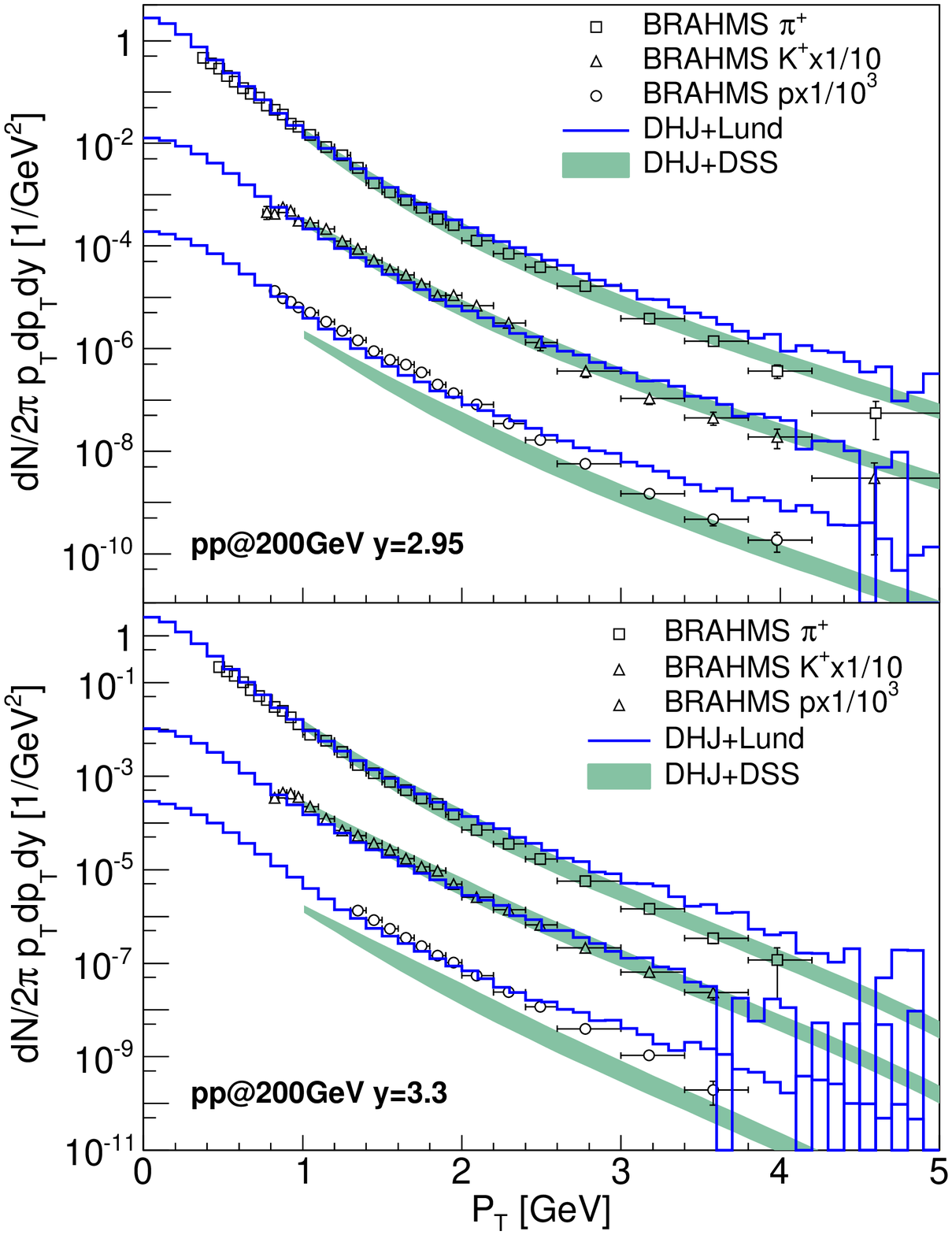}
\includegraphics[width=0.45\textwidth]{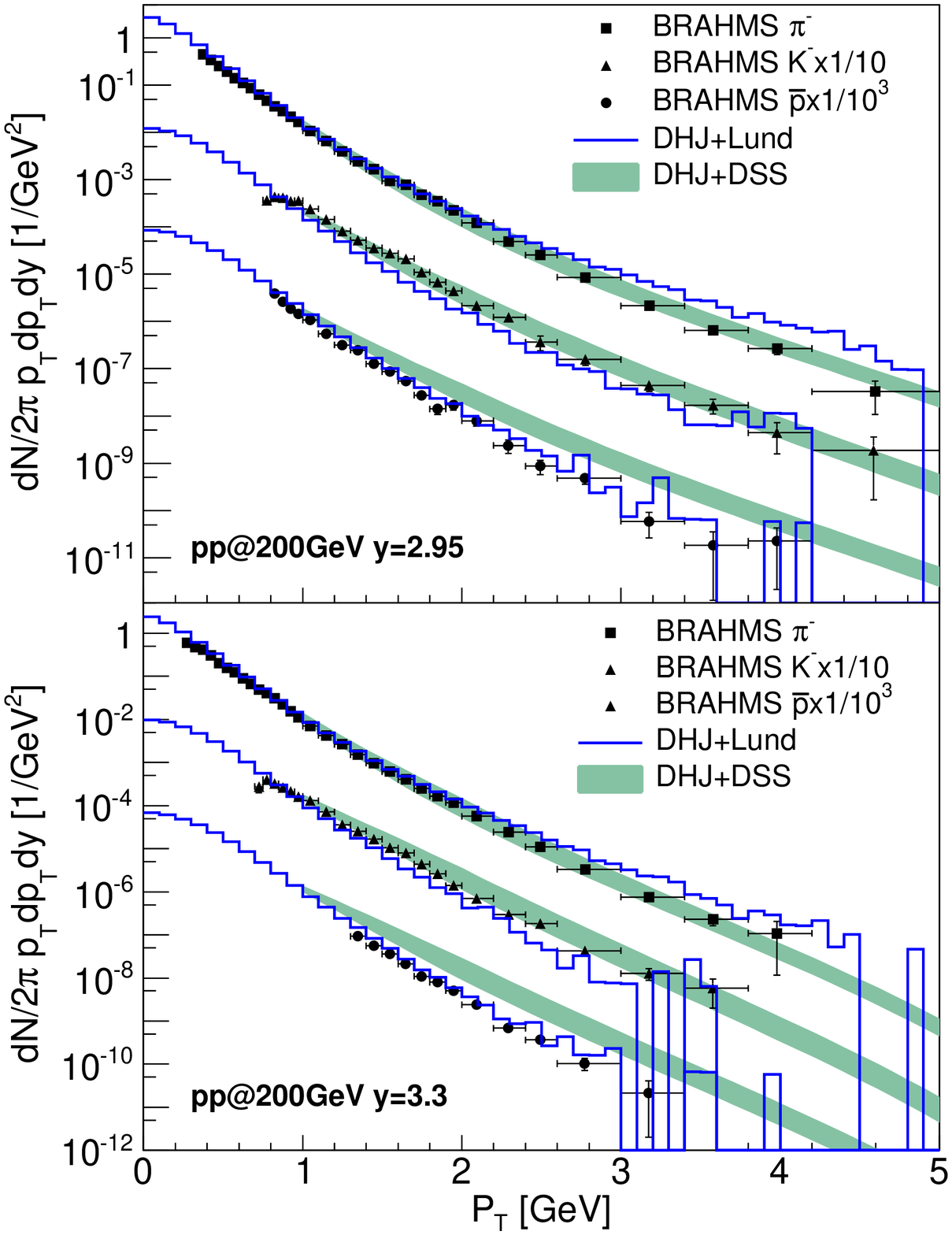}
\caption{Inclusive pions, kaons, protons and anti-protons
transverse momentum distributions in pp collisions at $\sqrt{s}=200$ GeV
at rapidities $y=2.95$ and $y=3.3$
obtained by the DHJ+Lund model (solid line) and the DHJ+DSS (box).
Positively (negatively) charged particles are shown in the left (right) panel.
Data are from BRAHMS collaboration~\cite{Arsene:2007jd}.
In the DHJ+DSS calculations, $K=2.5$ is used, and the width of the bands
indicates the different choice of the factorization scales within
$Q=(p_T/2,p_T)$.
\label{fig:pkp}}
\end{center}
\end{figure*}

In Fig.~\ref{fig:lphd200}, transverse
momentum distributions for negatively charged hadrons in pp collisions
at $\sqrt{s}=200$ GeV at rapidity 2.2 and 2.3 from the DHJ+LPHD model
are shown together with the experimental data from
BRAHMS~\cite{Arsene:2004ux}.
As an overall normalization, we used $K$-factor of 2.5.
We have checked that the results in Fig.~\ref{fig:lphd200}
is not sensitive to the value of the mass of the mini-jet $m_\text{jet}$
in the range from $m_\text{jet}=0.0$ to 0.5 GeV.
It is remarkable that under the assumption of LPHD the
DHJ framework works in low momentum region
for the description of transverse momentum distributions
even though it is not legitimate to use PDF in such a low $p_T$ range.
However, the LPHD approach does not work in high momentum region.
Instead, the fragmentation of the produced parton into hadrons
will be the appropriate picture there.
Indeed, the results of the DHJ formula
convoluted with the DSS fragmentation function~\cite{deFlorian:2007aj}
describe the data in high momentum region
as shown in the Fig.~\ref{fig:lphd200}.
The factorization scale dependence is also checked, and 
it is found that
the results for $Q=p_T/2$ is twice as large as the one for $Q=p_T$
with almost the same $p_T$ slope, where $K=2.5$ is used in the plot.
Thus scale dependence can be absorbed by changing $K$ factor. 
See also
Refs.~\cite{Albacete:2010bs,Fujii:2011fh,Fujii:2012zza,Albacete:2012xq,%
Lappi:2013zma}
for the CGC predictions for
forward hadron productions with fragmentation functions.

We now turn to the Monte-Carlo event generator, DHJ+Lund, results.
In Fig.~\ref{fig:dhj200} shown is the
negatively charged  hadron spectra and neutral pion
in pp collisions at $\sqrt{s}=200$ GeV obtained
by the DHJ+Lund model, together with
the BRAHMS data~\cite{Arsene:2004ux} and PHENIX data~\cite{Adare:2011sc}.
In the DHJ+Lund approach, hadron spectra from low to high momentum region
can be described within a single framework.
Note that we do not need intrinsic $k_{T}$ to fit the data,
which is often introduced in the conventional pQCD based models.
In the DHJ+LPHD approach, the spectrum in low momentum region is 
entirely described by the soft gluons, but in the DHJ+Lund approach,
it is modeled by the string fragmentation.

It would be informative to see the results from a model
based on the conventional collinear factorization of pQCD. 
For this purpose, we plot HIJING results in Fig.~\ref{fig:dhj200}
which we call HIJING2D, since it is a modified version
from the original HIJING2.0~\cite{Deng:2010mv}.
In HIJING, the nucleon-nucleon inelastic cross-section is given by
the eikonal formalism:
\begin{equation}
 \sigma_\text{in}= \int^{\infty}_0 d^2b[1-e^{2\chi(b,s)}],
\end{equation}
where the eikonal function $\chi(b,s)$ at an impact parameter $b$
and at the invariant mass $s$ is obtained as the sum of
the pQCD $2\to2$ cross-sections and the soft parton-parton collisions,
\begin{equation}
 \chi(b,s)=\frac{1}{2}[\sigma_\text{jet}T(b) + \sigma_\text{soft}T(b)]
,
\end{equation}
where $T(b)$ is the nucleon-nucleon overlap function and taken to be
(the Fourier transform of) the dipole form factor.
The LO pQCD $2\to2$ jet cross-section in pp collisions
at $\sqrt{s}=200$ GeV
is calculated 
with CTEQ6M PDF~\cite{cteq6}
and $K$-factor of 2.5,
assuming the $p_T$ cut-off $p_0=2.1$ GeV.
In HIJING2D, $\sigma_\text{soft}=58 \mb$ is used to fit the total pp 
cross-section and pseudo-rapidity distribution for charged hadrons
in pp collisions at $\sqrt{s}=200$ GeV.
The original HIJING model switches off the option for
the popcorn model~\cite{Eden97} in the Lund string fragmentation,
and the leading diquark does not break there.
But it is switched on in HIJING2D. 
HIJING2D results, shown in Fig.~\ref{fig:dhj200}, also describes
the experimental data very well.
This means that
we do not see any distinct effects of CGC in the charged
hadron data of BRAHMS and neural pion data of PHENIX in pp collisions
at $\sqrt{s}=200$ GeV.
This fact is actually encouraging: 
If one wants to construct a model in which both CGC
and pQCD processes are included, it is expected that 
the transition from pQCD
to CGC description will be smooth.

We also compare the neutral pion momentum distributions
in pp collisions at forward rapidities
from the STAR experiments~\cite{Adams:2006uz} in Fig.~\ref{fig:starpi0}.
The DHJ+Lund model overestimates the data by a factor of 2, 
although the slope is close to the data. 
We note that DHJ+Lund is consistent with the DHJ+DSS approach ($K=2.5$).
On the other hand, HIJING2D overestimates all data and its slopes are
much harder than the data.
In low momentum region $p_T<1$ GeV, both DHJ+Lund and HIJING2D 
model predictions are consistent with each other for all rapidities.
It is seen that there is a significant difference between HIJING2D
and DHJ+Lund approaches at momenta larger than 1 or 2 GeV.
The slopes in the HIJING2D results are much harder than those of
the DHJ+Lund model results which is more clearly seen at larger rapidity.
We have checked that PYTHIA6 gives quit similar results as HIJING2D,
but PYTHIA8.1 results are very close to the DHJ+Lund model results
as shown in Fig.~\ref{fig:starpi0}.
In Ref.~\cite{Adams:2006uz}, 
it was reported that NLO pQCD calculations
agree with the forward neutral pion from pp collisions,
despite that its spectrum is not consistent with the data
for $d$+Au collisions.

\subsection{Identified hadrons}

We turn now to the comparison of identified hadron spectra at forward
rapidities.
In Fig.~\ref{fig:pkp}, transverse momentum distributions
for pion, kaons, protons, and anti-protons from pp collisions
at $\sqrt{s}=200$ GeV at rapidities $y=2.95$ and $y=3.3$~\cite{Arsene:2007jd}
are compared with the DHJ+Lund model results.
Our model reasonably describes the BRAHMS data in
low momentum region $p_T<2$ GeV, but slopes are flatter at high $p_T$.
It should be noted that the DHJ+Lund model 
describes the proton and anti-proton yields simultaneously.
DHJ+DSS results for $Q=p_{T}$ and $Q=p_{T}/2$
with $K=2.5$ are also shown in Fig.~\ref{fig:pkp}.
The agreement with the BRAHMS data is reasonably good for pions and kaons.
However, the proton yield in DHJ+DSS is below the data
by a factor of two and five for rapidities $y=2.95$ and $y=3.3$, respectively. 
On the other hand, anti-proton yields are bigger
by a factor of two in DHD+DSS.

\begin{figure}[tpb]
\centerline{\includegraphics[width=0.45\textwidth]{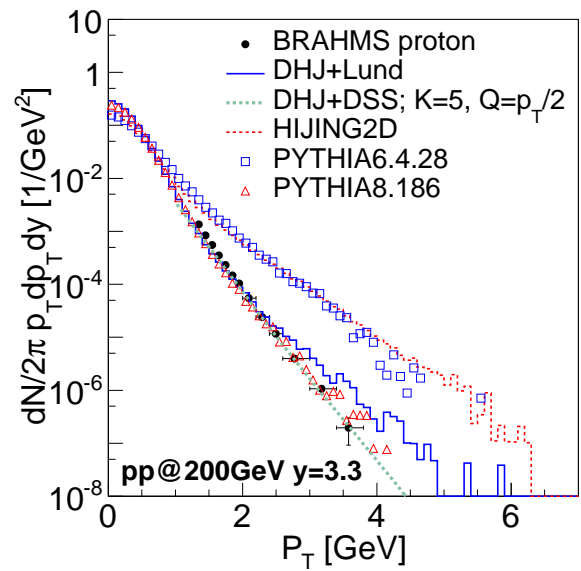}}
\caption{Inclusive proton invariant cross sections in pp collisions
at $\sqrt{s}=200$ GeV at $y=3.3$
are compared to the DHJ+LUND model (solid lines), HIJING2D (dashed lines),
DHJ+DSS (dotted lines), PYTHIA6.4.28 (squares), and
PYTHIA8.168 (triangles).
\label{fig:proton33}
}
\end{figure}

The DHJ+FF approach was applied to compute the proton 
yield in d+Au collisions at $\sqrt{s}=200$ GeV
in Ref.~\cite{Hayashigaki:2006ek}.
Fragmentation function extracted from the Lund string model~\cite{Wang1998}
scaled by AKK fragmentation function~\cite{Albino:2005me} 
was used in the calculations.
It is pointed out that the contributions from diquark fragmentation
play an essential role in the forward baryon production.
Note that both the DHJ+Lund model and HIJING2D take into account
these diquark contributions.
Thus, we expect that proton spectra in the DHJ+DSS approach 
also can be improved by taking such effects into account.

Various model predictions for the proton distribution at $y=3.3$ are
compared in Fig.~\ref{fig:proton33}.
DHJ+DSS with $K=5$ and $Q=p_{T}/2$ reproduces the proton data
very well, although this value of $K$-factor is inconsistent
with pion and kaon data.
We find that HIJING2D yields the same result as PYTHIA6, 
and both show much flatter proton distribution than the data.
On the other hand, the slope of the DHJ+Lund model at $p_T<3$ GeV
is close to the data,
although high $p_{T}$ part is flatter than the data, which
is not seen in the results with the DSS fragmentation function.
It is seen that PYTHIA8.1 result is in quite good agreement with
experimental data as shown in Fig.~\ref{fig:proton33}.
Thus, it is hard to see the distinct CGC effects
from the proton transverse momentum distribution 
in pp collisions at $\sqrt{s}=200$ GeV.

Finally, we should remark on the NLO pQCD results.
In Ref.~\cite{Arsene:2007jd}, it is reported that NLO pQCD calculations
describe the distribution of pions and kaons, but fail to fit the 
proton and anti-proton yields at the same time. 

\subsection{LHCf data}

\begin{figure*}[t]
\centerline{\includegraphics[width=0.9\textwidth]{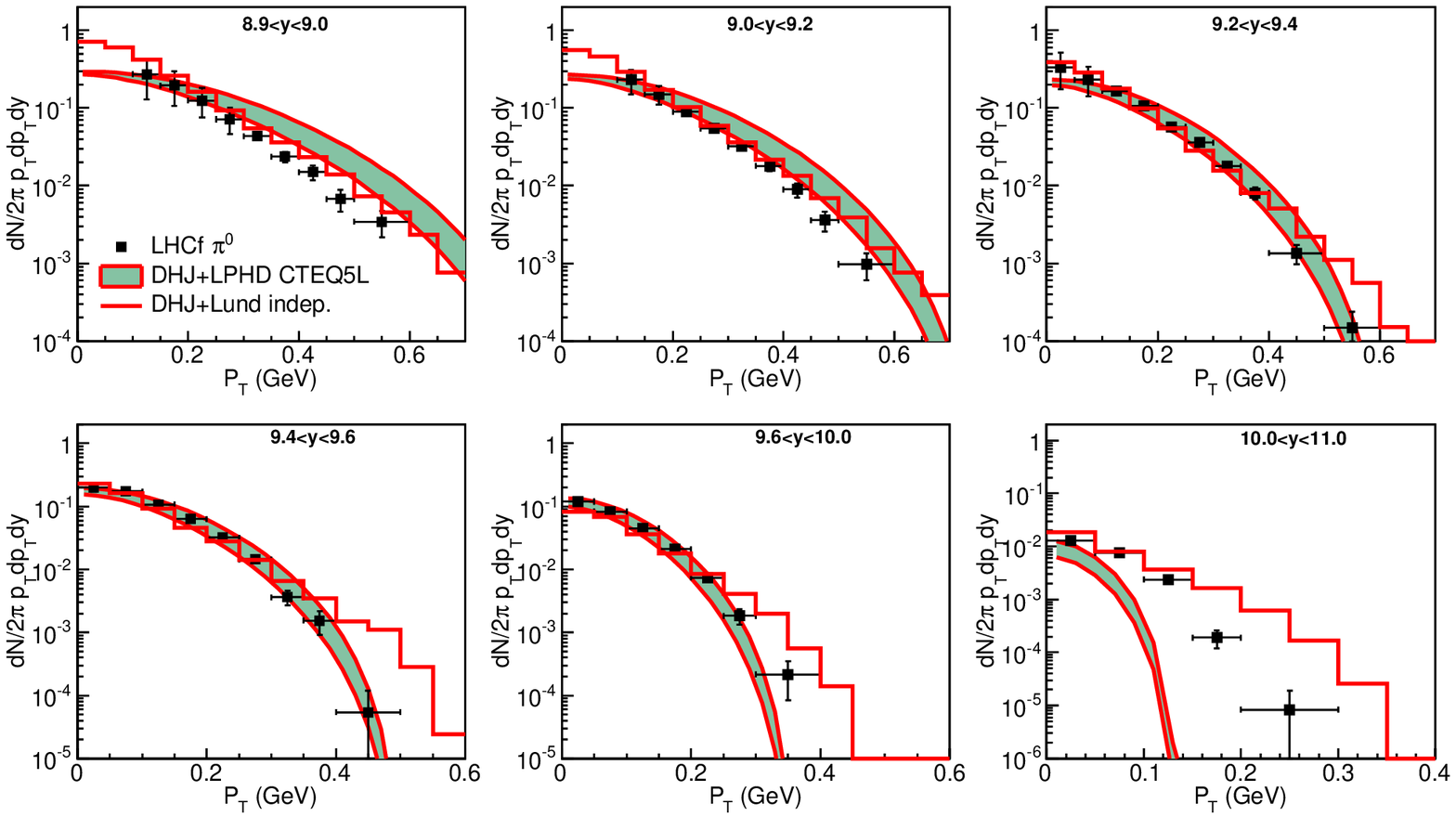}} \caption{
Comparison of invariant cross sections for neutral pion from LHCf in pp
collisions at $\sqrt{s}=7$ TeV with the DHJ+LPHD model and the DHJ with
independent fragmentation model of Lund.  }
\label{fig:dhjlphd} 
\centerline{\includegraphics[width=0.95\textwidth]{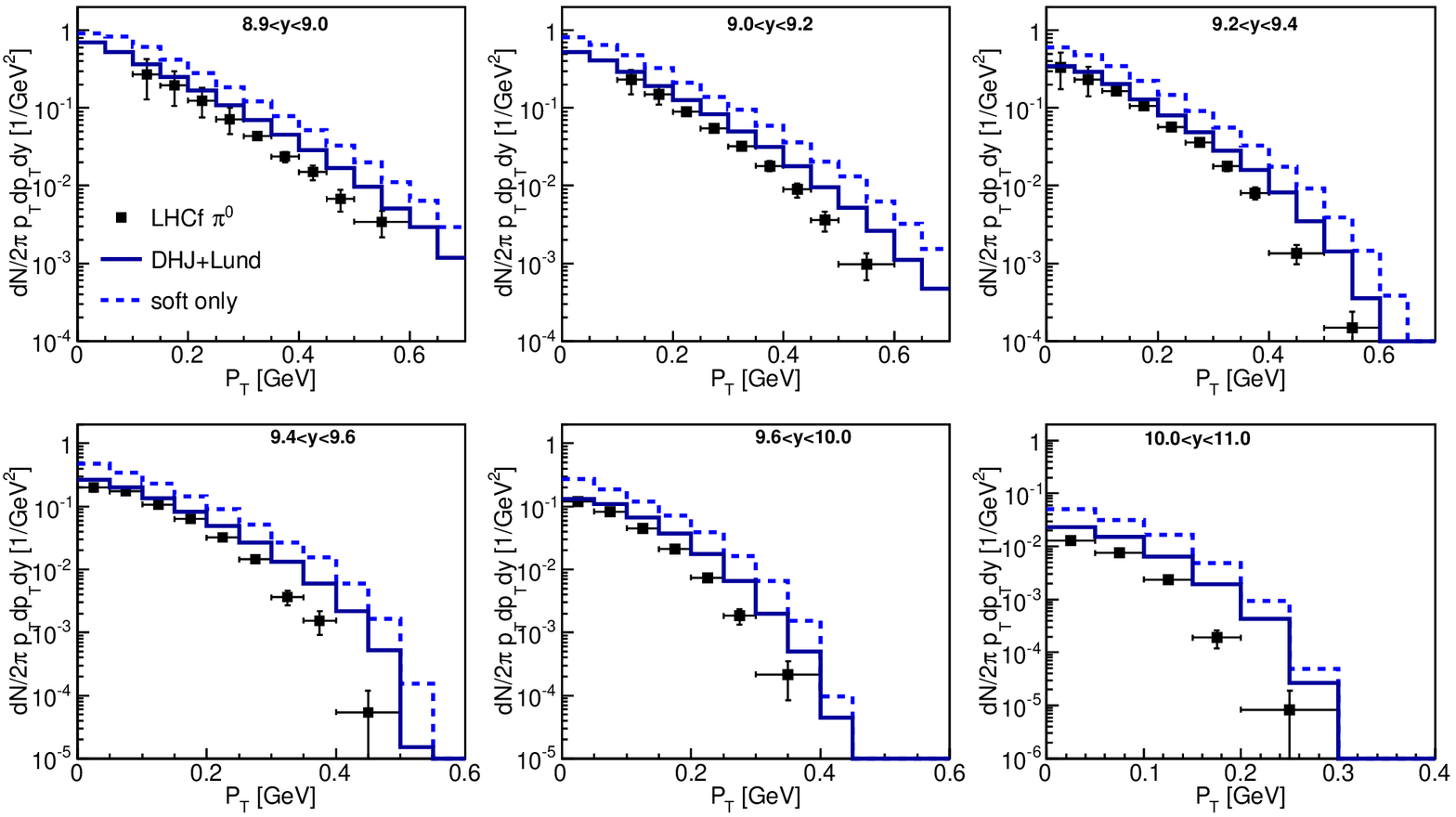}}
\caption{
Comparison of invariant cross sections for neutral pion
in pp collisions at $\sqrt{s}=7$ TeV from LHCf 
experiment~\cite{Adriani:2012ap}
with the DHJ+Lund model (solid lines). 
The results which only includes the soft string excitations
are shown by the dashed lines.
} \label{fig:dhjlund}
\end{figure*}

In this section, we analyze neutral pion transverse spectra
from LHCf experiment~\cite{Adriani:2012ap} in pp collisions
at $\sqrt{s}=7$ TeV.
Since measurement was performed in the very forward rapidity range
$8.9< y < 11.0$, the produced particles come from extremely high
$x_1$ and low $x_2$ regions of PDF.
In Ref.~\cite{Goncalves:2012bn}, LHCf data was analyzed based
on the DHJ formula convoluted with fragmentation function, 
and it was found that slopes are much steeper than the LHCf data.

First let us study the contributions of gluons and quarks from CGC
to the LHCf data.
In Fig.~\ref{fig:dhjlphd}
we present the DHJ+LPHD results of the neutral pion transverse
momentum spectra in pp collisions at $\sqrt{s}=7$ TeV
at different rapidities compared to LHCf data~\cite{Adriani:2012ap}.
The factorization scale of PDF is chosen to be 
$Q=\max(Q_s,p_T)$, where $Q_s$ is a saturation scale and
$p_T$ is the transverse momentum of a parton.
Sensitivities to the choice of the factorization scale of the PDF
is also shown in the Fig.~\ref{fig:dhjlphd} 
as a band between the results with $Q/2$ and $2Q$.
We have checked the sensitivity of the particle spectra to the PDF.
The calculations with GRV94L and CTEQ6M PDF
yield results similar to the CTEQ5L results.
DHJ+LPHD describes the correct slopes for most of the rapidity bins.
However, we found that in order to fit the data we need 
an extremely large $K=22~ (15)$ factor 
assuming the inelastic (non-diffractive)
cross section of $\sigma_\text{inel}=73.6$ mb~\cite{Adriani:2012ap}
($\sigma_\text{nondif}=48.45$ mb).
(If we do not include the factor $\sigma_0/2$, $K$ factor will be $K=5$).
We conclude, therefore, that DHJ+LPHD approach does not work for explaining
the LHCf forward pion data.

We also test a different approach in which gluons from the DHJ
formula  fragment into hadrons independently.
The results from the independent fragmentation 
are plotted by histogram in Fig.~\ref{fig:dhjlphd}.
For this hadronization process, the Lund model is also used.
In this model, reasonable $K$ factor $K=2.5$ is obtained to
fit the data.
It is interesting to see that the Lund independent fragmentation model
describes the correct slopes.
However, independent fragmentation picture may not be adequate 
to describe very low momentum hadrons.

Finally, the results of the DHJ+Lund model for neutral pion
at $\sqrt{s}=7$ TeV are
shown in Fig.~\ref{fig:dhjlund} together with the results in which
only soft interactions are included.
By the soft interaction, we mean that
two strings are
excited according to the formula~(\ref{eq:dpm}), and possible
additional string productions between sea-quarks are neglected.
Increase of the yield in the model with the soft interaction only
may be understood as follows: In this model, strings are
stretched almost parallel to the beam direction and likely to 
produce pions in the forward region. On the other hand, 
forward pion production is suppressed by the gluons attached in the string,
when CGC gluons are included.
The DHJ+Lund model agrees with the data at low momentum, 
and its slopes are slightly harder than the LHCf data.
The effects of diffractive interactions are checked by PYTHIA6,
and we found that single and double diffractive interactions yield
similar results as non-diffractive one, except the single-diffractive
scattering in which one of forward particle goes to the negative rapidity
direction
whose contribution is estimated about 12\% of inelastic cross section.
From our analysis,  most of the neutral pions in the ultra-forward region 
can be explained by the soft physics which come from the decay of strings.

As pointed out in Ref.~\cite{Adriani:2012ap},
pion distribution is sensitive to the choice of baryon production model.
Indeed, we have checked that if the popcorn model~\cite{Eden97} is switched off,
much steeper spectra is obtained which is inconsistent with the data.

It is reported from LHCf~\cite{Adriani:2014mfa} that
nuclear modification factor of forward neutral pion 
in proton-lead collisions at $\sqrt{s_{NN}}=5.02$ TeV
exhibits strong suppression, and it increases with transverse momentum.
However, hadronic interaction model predictions
show almost flat $p_T$ dependence.
It is interesting to explore p-Pb collisions  within our approach
in the future.

\section{Summary}
\label{sec:summary}

In this paper, we have developed a Monte Carlo version
of the DHJ formula (DHJ+Lund model) for proton-proton collisions.
In this model, we explicitly generate $gg\to g$ and $gq\to q$
scatterings together with the initial and final state radiations.
Those produced partons are connected with the remnant excited strings,
and decay into hadrons by the Lund string fragmentation model.

We have compared the results of our model with hadron transverse 
momentum distributions
at forward rapidities and from low to high momentum regions
at both RHIC and LHC energies.
We also studied the spectra for identified hadrons within our model.
The model provides a unified description of the hadron spectra
from non-perturbative low to high momentum region.
It is shown that the DHJ+Lund  model yields consistent results 
with the approach
with fragmentation function for charged, pion, and kaon spectra
in high $p_T$ regions.
Some improved description of the
baryon production at forward rapidities was seen.
We also analyzed LHCf ultra-forward neutral pion spectra by both
the DHJ+LPHD and DHJ+Lund models, and found that
dominant process for such ultra-forward pion is 
the soft excitation of strings and their fragmentations.
As a future work, we are planning to extend the model to
proton-nucleus and nucleus-nucleus collisions.

\acknowledgements
Y.N. is  thankful to Adrian Dumitru for useful comments.
The authors thank K. Kasahara, H. Menjo, and  T. Sako
for providing the main program for PYTHIA8.
This work was partially supported
by a Grant-in-Aid for Scientific Research (B) (22340064).


\begin{thebibliography}{99}

\bibitem{Heinz:2013th} 
U.~Heinz and R.~Snellings,
Ann.\ Rev.\ Nucl.\ Part.\ Sci.\  {\bf 63}, 123 (2013)
[arXiv:1301.2826 [nucl-th]].

\bibitem{Engel:2011zzb} R.~Engel, D.~Heck and T.~Pierog,
Ann.\ Rev.\ Nucl.\ Part.\ Sci.\ {\bf 61} (2011) 467.

\bibitem{MV}
L.D. McLerran, R. Venugopalan, Phys. Rev. D 49 (1994) 2233;
Phys. Rev. D 49 (1994) 3352; Phys. Rev. D 50 (1994) 2225.


\bibitem{GIJV}
F. Gelis, E. Iancu, J. Jalilian-Marian, R. Venugopalan, 
Ann. Rev. Nucl. Part. Sci. {\bf 60} (2010) 463-489.


\bibitem{KV}
A. Krasnitz, R. Venugopalan, Nucl. Phys. B 557 (1999) 237;
A. Krasnitz, Y. Nara, R. Venugopalan, Phys. Rev. Lett. 87 (2001) 192302;
Krasnitz, Y. Nara, R. Venugopalan, Nucl. Phys. A 727 (2003) 427.

\bibitem{Lappi} T. Lappi, Phys. Rev. C 67 (2003) 054903;
Eur. Phys. J. C 55 (2008) 285;
Phys. Lett. B 703 (2011) 325.

\bibitem{IPGlasma}
B.~Schenke, P.~Tribedy and R.~Venugopalan,
Phys.\ Rev.\ Lett.\  {\bf 108}, 252301 (2012)
[arXiv:1202.6646 [nucl-th]];
B.~Schenke, P.~Tribedy and R.~Venugopalan,
Phys.\ Rev.\ C {\bf 86}, 034908 (2012)
[arXiv:1206.6805 [hep-ph]];
C.~Gale, S.~Jeon, B.~Schenke, P.~Tribedy and R.~Venugopalan,
Phys.\ Rev.\ Lett.\  {\bf 110}, 012302 (2013)
[arXiv:1209.6330 [nucl-th]];
B.~Schenke, P.~Tribedy and R.~Venugopalan,
Phys.\ Rev.\ C {\bf 89}, 024901 (2014)
[arXiv:1311.3636 [hep-ph]].


\bibitem{ktfact}
Y.~V.~Kovchegov and K.~Tuchin,
Phys.\ Rev.\ D {\bf 65}, 074026 (2002)
[hep-ph/0111362].


\bibitem{KLN}
D. Kharzeev and M. Nardi, Phys. Lett. {\bf B} 507, 121 (2001);
D. Kharzeev and E. Levin, Phys. Lett. {\bf B} 523, 79 (2001);
D. Kharzeev, E. Levin, and M. Nardi, Nucl. Phys. A 730, 448 (2004);
D. Kharzeev, E. Levin, and M. Nardi, Phys. Rev. {\bf C} 71, 054903 (2005);
D. Kharzeev, E. Levin, and M. Nardi, Nucl. Phys. {\bf A 747}, 609 (2005).
A. Dumitru, D.E. Kharzeev, E.M. Levin, Y. Nara, Phys. Rev. {\bf C 85} 044920,
(2012).

\bibitem{fKLN}
H.~J.~Drescher, A.~Dumitru, A.~Hayashigaki and Y.~Nara,
Phys.\ Rev.\ C {\bf 74}, 044905 (2006)
[nucl-th/0605012].



\bibitem{Levin:2010zy} 
  E.~Levin and A.~H.~Rezaeian,
      Phys.\ Rev.\ D {\bf 82}, 054003 (2010)
        [arXiv:1007.2430 [hep-ph]].

\bibitem{Levin:2010dw} 
E.~Levin and A.~H.~Rezaeian,
Phys.\ Rev.\ D {\bf 82}, 014022 (2010)
[arXiv:1005.0631 [hep-ph]].

\bibitem{Levin:2011hr} 
  E.~Levin and A.~H.~Rezaeian,
      Phys.\ Rev.\ D {\bf 83}, 114001 (2011)
        [arXiv:1102.2385 [hep-ph]].

\bibitem{LPHD}
Ya.I. Azimov, Yu.L. Dokshitzer, V.A. Khoze and S.I. Troyan, 
Z. Phys.  {\bf C 27}, 65 (1985); ibid, {bf 31}, 213 (1986).

\bibitem{Kharzeev:2004yx} 
D.~Kharzeev, Y.~V.~Kovchegov and K.~Tuchin,
Phys.\ Lett.\ B {\bf 599}, 23 (2004)
[hep-ph/0405045].


\bibitem{Albacete:2010bs} 
J.~L.~Albacete and C.~Marquet,
Phys.\ Lett.\ B {\bf 687}, 174 (2010)
[arXiv:1001.1378 [hep-ph]].

\bibitem{Fujii:2011fh} 
  H.~Fujii, K.~Itakura, Y.~Kitadono and Y.~Nara,
      J.\ Phys.\ G {\bf 38}, 124125 (2011)
        [arXiv:1107.1333 [hep-ph]].

\bibitem{Fujii:2012zza} 
  H.~Fujii, K.~Itakura and Y.~Nara,
      Prog.\ Theor.\ Phys.\ Suppl.\  {\bf 193}, 216 (2012).


\bibitem{Albacete:2012xq} 
J.~L.~Albacete, A.~Dumitru, H.~Fujii and Y.~Nara,
Nucl.\ Phys.\ A {\bf 897}, 1 (2013)
[arXiv:1209.2001 [hep-ph]].




\bibitem{Lappi:2013zma} 
T.~Lappi and H.~M\"antysaari,
Phys.\ Rev.\ D {\bf 88}, 114020 (2013)
[arXiv:1309.6963 [hep-ph]].



\bibitem{DHJ}
A.~Dumitru, A.~Hayashigaki and J.~Jalilian-Marian,
Nucl.\ Phys.\ A {\bf 765}, 464 (2006) [hep-ph/0506308];
Nucl.\ Phys.\ A {\bf 770}, 57 (2006)
[hep-ph/0512129].

\bibitem{Hayashigaki:2006ek} 
A.~Hayashigaki,
Nucl.\ Phys.\ A {\bf 775}, 51 (2006)
[hep-ph/0604173].

\bibitem{MehtarTani:2009dv} 
Y.~Mehtar-Tani and G.~Wolschin,
Phys.\ Rev.\ C {\bf 80}, 054905 (2009)
[arXiv:0907.5444 [hep-ph]].

\bibitem{Duraes:2014jxa} 
  F.~O.~Duraes, A.~V.~Giannini, V.~P.~Goncalves and F.~S.~Navarra,
      Phys.\ Rev.\ C {\bf 89}, 035205 (2014)
        [arXiv:1401.7888 [hep-ph]].



\bibitem{rcBKkern}
I.~Balitsky,
  Phys.\ Rev.\ {\bf D} {\bf 75}, 014001 (2007);
  Y.~V.~Kovchegov and H.~Weigert,
  Nucl.\ Phys.\ A {\bf 784}, 188 (2007);
  Nucl.\ Phys.\ A {\bf 789}, 260 (2007).

\bibitem{AK}
  J.~L.~Albacete and Y.~V.~Kovchegov,
  Phys.\ Rev.\ {\bf D} {\bf 75}, 125021 (2007).

\bibitem{AAMQS}
J.~L.~Albacete, N.~Armesto, J.~G.~Milhano, P.~Quiroga-Arias and C.~A.~Salgado,
Eur.\ Phys.\ J.\ C71, 1705 (2011); see, also,
J.~L.~Albacete, N.~Armesto, J.~G.~Milhano and C.~A.~Salgado,
Phys.\ Rev.\ D {\bf 80}, 034031 (2009).

\bibitem{Albacete:2013tpa} 
  J.~L.~Albacete, A.~Dumitru and C.~Marquet,
      Int.\ J.\ Mod.\ Phys.\ A {\bf 28}, 1340010 (2013)
        [arXiv:1302.6433 [hep-ph]].


\bibitem{NLODHJ}
G. A. Chirilli, B. -W. Xiao and F. Yuan,
Phys. Rev. Lett. {\bf 108}, 122301 (2012);
Phys. Rev. {\bf D 86}, 054005 (2012);
A. M. Stasto, B. -W. Xiao and D. Zaslavsky, 
Phys. Rev. Lett. {\bf 112}, 012302 (2014);
A.~M.~Sta\'sto, B.~W.~Xiao, F.~Yuan and D.~Zaslavsky,
Phys.\ Rev.\ D {\bf 90}, 014047 (2014)
[arXiv:1405.6311 [hep-ph]].


\bibitem{MCKLN} H.-J. Drescher and Y. Nara, Phys. Rev. C 75, 034905 (2007);
76, 041903(R) (2007).

\bibitem{MCKT}
J.~L.~ALbacete and A.~Dumitru,
arXiv:1011.5161 [hep-ph];
J.~L.~Albacete, A.~Dumitru and Y.~Nara,
J.\ Phys.\ Conf.\ Ser.\  {\bf 316}, 012011 (2011)
[arXiv:1106.0978 [nucl-th]].

\bibitem{Hirano:2012kj}
T.~Hirano, P.~Huovinen, K.~Murase and Y.~Nara,
Prog.\ Part.\ Nucl.\ Phys.\  {\bf 70}, 108 (2013)
[arXiv:1204.5814 [nucl-th]].
T.~Hirano and Y.~Nara,
PTEP {\bf 2012}, 01A203 (2012)
[arXiv:1203.4418 [nucl-th]].

\bibitem{Hirano:2010jg} 
T.~Hirano, P.~Huovinen and Y.~Nara,
Phys.\ Rev.\ C {\bf 83}, 021902 (2011);
Phys.\ Rev.\ C {\bf 84}, 011901 (2011).


\bibitem{Qiu:2011hf} 
Z.~Qiu, C.~Shen and U.~Heinz,
Phys.\ Lett.\ B {\bf 707}, 151 (2012);
U.~Heinz, Z.~Qiu and C.~Shen,
Phys.\ Rev.\ C {\bf 87}, no. 3, 034913 (2013);
H.~Song, S.~Bass and U.~W.~Heinz,
Phys.\ Rev.\ C {\bf 89}, 034919 (2014).



\bibitem{Drescher:2004sd} 
H.~J.~Drescher, A.~Dumitru and M.~Strikman,
Phys.\ Rev.\ Lett.\  {\bf 94}, 231801 (2005);
  hep-ph/0501165.


\bibitem{Lai:1999wy} 
H.~L.~Lai {\it et al.}  [CTEQ Collaboration],
Eur.\ Phys.\ J.\ C {\bf 12}, 375 (2000)
[hep-ph/9903282].


\bibitem{pythia6}
T.~Sjostrand, S.~Mrenna and P.~Z.~Skands,
JHEP {\bf 0605} (2006), 026.

\bibitem{pythia} T. Sj{\" o}strand,
Comp. Phys. Comm. {\bf 82} (1994), 74;\\
    \url{http://www.thep.lu.se/tf2/staff/torbjorn/Pythia.html}.

\bibitem{pythia8}
T.~Sjostrand, S.~Mrenna and P.~Z.~Skands,
Comput.\ Phys.\ Commun.\  {\bf 178}, 852 (2008)
[arXiv:0710.3820 [hep-ph]].

\bibitem{Eikonal} J.M. Butterworth, J.R. Forshaw, M.H. Seymour,
Z. Phys. {\bf C} 72, 637 (1996), hep-ph/9601371;
L. Durand, P. Hong, Phys. Rev. Lett. {\bf 58}, 303 (1987);
L. Durand, H. Pi, Phys. Rev. {\bf D} 40, 1436 (1989).


\bibitem{Wang:1991hta} 
  X.~N.~Wang and M.~Gyulassy,
      Phys.\ Rev.\ D {\bf 44}, 3501 (1991).
\bibitem{Gyulassy:1994ew} 
  M.~Gyulassy and X.~N.~Wang,
      Comput.\ Phys.\ Commun.\  {\bf 83}, 307 (1994)
        [nucl-th/9502021].

\bibitem{Deng:2010mv} 
W.~T.~Deng, X.~N.~Wang and R.~Xu,
Phys.\ Rev.\ C {\bf 83}, 014915 (2011)
[arXiv:1008.1841 [hep-ph]].

\bibitem{Fletcher:1994bd} 
  R.~S.~Fletcher, T.~K.~Gaisser, P.~Lipari and T.~Stanev,
      Phys.\ Rev.\ D {\bf 50}, 5710 (1994).

\bibitem{Ahn:2009wx} 
  E.~J.~Ahn, R.~Engel, T.~K.~Gaisser, P.~Lipari and T.~Stanev,
      Phys.\ Rev.\ D {\bf 80}, 094003 (2009)
        [arXiv:0906.4113 [hep-ph]].

\bibitem{Bahr:2008dy} 
  M.~Bahr, S.~Gieseke and M.~H.~Seymour,
      JHEP {\bf 0807}, 076 (2008)
        [arXiv:0803.3633 [hep-ph]].

\bibitem{Bahr:2008pv} 
  M.~Bahr, S.~Gieseke, M.~A.~Gigg, D.~Grellscheid, K.~Hamilton,
  O.~Latunde-Dada, S.~Platzer and P.~Richardson {\it et al.},
      Eur.\ Phys.\ J.\ C {\bf 58}, 639 (2008)
        [arXiv:0803.0883 [hep-ph]].





\bibitem{Andersson:1983ia} 
B.~Andersson, G.~Gustafson, G.~Ingelman and T.~Sjostrand,
Phys.\ Rept.\  {\bf 97}, 31 (1983);
Bo Andersson, \textit{The Lund Model},
Cambridge University Press, 1998




\bibitem{popparam}
We set $\texttt{MSTJ(12)}=5$ for the choice of baryon production model.
According to the suggestion in Ref.\cite{pythia6},
the parameters $\texttt{PARJ(1)}=0.2$ for the suppresion of
diquark-antidiquark pair production
and $\texttt{PARJ(18)}=0.19$ are set.


\bibitem{Arsene:2004ux} 
I.~Arsene {\it et al.}  [BRAHMS Collaboration],
Phys.\ Rev.\ Lett.\  {\bf 93}, 242303 (2004)
[nucl-ex/0403005].


\bibitem{deFlorian:2007aj} 
D.~de Florian, R.~Sassot and M.~Stratmann,
Phys.\ Rev.\ D {\bf 75}, 114010 (2007)
[hep-ph/0703242 [HEP-PH]];
Phys.\ Rev.\ D {\bf 76}, 074033 (2007)
[arXiv:0707.1506 [hep-ph]].

\bibitem{Adare:2011sc} 
A.~Adare {\it et al.}  [PHENIX Collaboration],
Phys.\ Rev.\ Lett.\  {\bf 107}, 172301 (2011)
[arXiv:1105.5112 [nucl-ex]].


\bibitem{cteq6}
J.~Pumplin, D.~R.~Stump, J.~Huston, H.~L.~Lai, P.~M.~Nadolsky and W.~K.~Tung,
JHEP {\bf 0207}, 012 (2002)
[hep-ph/0201195].

\bibitem{Eden97}
P.~Eden and G.~Gustafson,
Z.\ Phys.\ C {\bf 75}, 41 (1997)
[hep-ph/9606454].


\bibitem{Adams:2006uz} 
J.~Adams {\it et al.}  [STAR Collaboration],
Phys.\ Rev.\ Lett.\  {\bf 97}, 152302 (2006)
[nucl-ex/0602011].




\bibitem{Arsene:2007jd} 
I.~Arsene {\it et al.}  [BRAHMS Collaboration],
Phys.\ Rev.\ Lett.\  {\bf 98}, 252001 (2007)
[hep-ex/0701041].




\bibitem{Wang1998}
X.-N. Wang, Phys. Rev. {\bf C58}, 2321 (1998).

\bibitem{Albino:2005me} 
 S.~Albino, B.~A.~Kniehl and G.~Kramer,
Nucl.\ Phys.\ B {\bf 725}, 181 (2005)
[hep-ph/0502188].


\bibitem{Adriani:2012ap} 
O.~Adriani {\it et al.}  [LHCf Collaboration],
Phys.\ Rev.\ D {\bf 86}, 092001 (2012)
[arXiv:1205.4578 [hep-ex]].


\bibitem{Goncalves:2012bn} 
V.~P.~Goncalves and M.~L.~L.~da Silva,
Nucl.\ Phys.\ A {\bf 906}, 28 (2013)
[arXiv:1210.6311 [hep-ph]].

\bibitem{Adriani:2014mfa} 
O.~Adriani, E.~Berti, L.~Bonechi, M.~Bongi, G.~Castellini, R.~D'Alessandro,
M.~Del Prete and M.~Haguenauer {\it et al.},
Phys.\ Rev.\ C {\bf 89}, 065209 (2014)
[arXiv:1403.7845 [nucl-ex]].


\end{thebibliography}

\end{document}